\documentclass[useAMS,usenatbib]{mn2e}

\usepackage{amsmath}
\usepackage{amssymb}
\usepackage{amsfonts}
\usepackage{amstext}
\usepackage{graphicx}
\usepackage{fancybox}
\usepackage[usenames,dvipsnames]{color}
\usepackage{pstricks}
\usepackage{bm}
\usepackage{soul, color}
\usepackage{latexsym}
\usepackage{pifont}
\usepackage{shorttoc}
\usepackage{subfigure}
\usepackage{textcomp} 
\usepackage{hyperref}
\usepackage{bm}
\usepackage{stfloats}
\usepackage[off]{auto-pst-pdf}

\newcommand{\lya}{Ly$\alpha$\ }
\newcommand{\kms}{\, {\rm km \, s}^{-1}}
\newcommand{\mpc}{\, {\rm Mpc}}
\newcommand{\hmpc}{\, h^{-1}\, {\rm Mpc}}

\begin{document}

\title[Radiation effect on \lya forest correlation]{On the effect of the ionising background on the \lya forest autocorrelation function}
\author[S. Gontcho A Gontcho, J. Miralda-Escud\'e, N. G. Busca]
{Satya ~Gontcho A Gontcho$^1$, Jordi ~Miralda-Escud\'e$^{1,2}$,
 \vspace{0.4cm}  Nicol\'as G. ~Busca$^{3,4,5}$ \\ 
$^1$Institut de Ci\`encies del Cosmos, Universitat de Barcelona/IEEC,
Barcelona 08028, Catalonia, Spain\\ 
$^2$Instituci\'o Catalana de Recerca i Estudis Avan\c{c}ats, Barcelona,
Catalonia, Spain\\
$^3$APC, Universit\'e Paris Diderot-Paris 7, CNRS/IN2P3, 10 rue A. Domon
\& L. Duquet, Paris, France\\
$^4$Observat\'orio Nacional, Rua Gal.~Jos\'e Cristino 77, Rio de Janeiro, RJ - 20921-400, Brazil\\
$^5$Laborat\'orio Interinstitucional de e-Astronomia, - LIneA, Rua Gal.Jos\'e Cristino 77, Rio de Janeiro, RJ - 20921-400, Brazil}

%\begin{document}

\date{Accepted 2014 April 29 ; Received 2014 April 22; in original form 2014 February 5}

\pagerange{\pageref{firstpage}--\pageref{lastpage}} \pubyear{2014}

\maketitle

\label{firstpage}

\begin{abstract}
 ~\par  An analytical framework is presented to understand the effects of a fluctuating
intensity of the cosmic ionising background on the correlations of the \lya
forest transmission fraction measured in quasar spectra. In the absence of
intensity fluctuations, the \lya power spectrum should have the expected cold
dark matter power spectrum with redshift distortions in the linear regime, with
a bias factor $b_\delta$ and a redshift distortion parameter $\beta$ that
depend on redshift but are independent of scale. The intensity fluctuations 
introduce a scale dependence in both $b_\delta$ and $\beta$, but keeping
their product $b_\delta \beta$ fixed. Observations of the \lya correlations
and cross-correlations with radiation sources like those being done at present
in the BOSS survey of SDSS-III \citep{Busca2013,Slosar2013,Font14} have the
potential to measure this scale dependence, which reflects the biasing
properties of the sources and absorbers of the ionising background. We also
compute a second term affecting the \lya spectrum, due to shot noise in the
sources of radiation. This term is very large if luminous quasars are assumed
to produce the ionising background and to emit isotropically with a constant
luminosity, but should be reduced by a contribution from galaxies, and by the
finite lifetime and anisotropic emission of quasars.
\end{abstract}


\section{Introduction}

 ~\par The \lya forest absorption measured in spectra of high-redshift
quasars has now been established as a powerful tracer of large-scale
structure. Assuming that the intrinsic continuum spectrum of the
observed quasar can be accurately modelled, then the observed flux
divided by the fitted continuum yields the transmitted fraction,
$F=e^{-\tau}$ (where $\tau$ is the optical depth), at every wavelength
pixel. This one-dimensional map that is obtained from the spectrum of
every observed source is related (neglecting the contamination by metal
lines) to the gas density, temperature and peculiar velocity of the
hydrogen gas in the intergalactic medium that is intercepted by the line
of sight.

 ~\par After the initial measurements of the \lya power spectrum along
the line of sight from individual spectra
\citep{Croft1998,Croft1999,McDonald2000,Croft2002,McDonald2006},
the first determination of the power spectrum of the \lya forest in
three-dimensional redshift space came with the BOSS survey of SDSS-III
\citep{Eisenstein2011,Dawson2013}. Analysis of the first
14000 quasars led to the detection of redshift space distortions
\citep{Slosar2011}, as expected in a simple biased linear theory where
the \lya power spectrum follows that of the dark matter with two bias
parameters, reflecting the large-scale variation of the mean \lya
transmission with the fluctuation in the mean mass density and peculiar
velocity gradient.

~\par However, large-scale fluctuations in the \lya forest can also be
affected by variations in the intensity of the ionising background
radiation, as well as the imprint that reionisation may have left on the
gas temperature distribution as a function of gas density. These
effects have been studied and discussed by several authors in the past.
Analytic models of randomly distributed sources were considered by
\cite{Zuo1992}, and numerical realizations of random sources to compute
the fluctuation properties of the ionising background
were used in several subsequent papers
\citep{Croft1999,Croft2004,Meiksin2004,McDonald2005,Ho09,White2010}.
The impact of these ionising background fluctuations on the \lya forest
were found to be generally small compared to the intrinsic \lya forest
fluctuations due to the large-scale structure of the mass distribution.
However, as pointed out in the early work of \cite{Croft1999},
the long mean free path of ionising radiation in the intergalactic
medium at $z \sim 3$ implies that the fluctuations induced by the
ionising background can become relatively more important in the limit of
very large scales. These large scales are now becoming highly relevant
with the recent detections of the BAO peak in the \lya forest
\citep{Busca2013,Slosar2013,Font14,Delubac2014}.

%can therefore modify
%the shape of the \lya power spectrum from that expected in the standard
%Cold Dark Matter with constant linear bias factors, which assumes a
%uniform ionising radiation intensity and temperature-density relation.

 ~\par In this paper we reanalyse with an analytic method the impact of
large-scale fluctuations in the ionising radiation intensity and the gas
temperature-density relation on the observable redshift space \lya power
spectrum. There are two independent effects on the power spectrum.
The first arises from the
clustering of sources and absorbers of radiation, which are assumed to
trace the large-scale mass density fluctuations, each with their own
bias factor. This clustering term is independent of the luminosity
function, variability and anisotropic emission of the sources, as well
as the size or other geometric properties of the absorbers: it depends
only on how the density of sources and absorbers follow the underlying
large-scale structure. The second effect is due to the fluctuations in
the radiation intensity that arises from shot noise in the number of
sources. This second term is independent of the source clustering, but
depends on other
source characteristics like the luminosity function. An analytical
framework to treat these contributions to the \lya power spectrum is
described in section 2, and results for simple illustrating models are
presented in section 3, with a discussion and conclusions in section 4.
We use a Cold Dark matter cosmological model with parameter values that
are consistent with the \cite{Planck}: $H_0=67.3$ $\kms\mpc^{-1}$,
baryon density $\Omega_{b}h^2=0.02205$, 
%cold dark matter density $\Omega_{c}h^2=0.1199$, 
$\Omega_m=0.315$, $n_s=0.96$ and $\sigma_8=0.856$. 

~\par As this paper was being finalized, we became aware of the work by
Pontzen (2014), presenting very similar ideas as here. We mention in
section 4 the similarities and differences between the two papers.

\section{Analytic Formalism}\label{sec:nm}

~\par The use of the \lya forest as a tracer of large-scale structure
lies on the principle that, when averaged over a large scale, the mean
value of the transmission fluctuation through the \lya forest,
$\delta_\alpha=F/\bar F(z)-1$ (where $F$ is equal to the observed flux
divided by a model quasar continuum, and $\bar F$ is the mean value of
$F$ over all the universe at redshift $z$), has a linear relation to the
local deformation tensor of large-scale structure when smoothed in the
same way over a large scale. On small scales, the distribution of
$\delta_\alpha$ and its correlations have a complex dependence on the
physics of non-linear collapse of the intergalactic gas into filaments
and halos, and the shock-heating, ionisation and cooling of the gas.
However, on large scales all these effects are absorbed into a first
order dependence of $\delta_\alpha$ on the local
deformation tensor in the linear regime \citep[e.g.,][]{Kaiser1987},
\begin{equation}
 {1\over H(z)}{\partial v_i\over \partial x_j} ~,
\end{equation}
where ${\textbf v}$ is the peculiar velocity smoothed over a large scale
in the same way as $F$, $x$ is the comoving coordinate, and $H(z)$ is
the Hubble constant at redshift $z$. For an observer measuring $F$ along
a direction specified by a unit vector ${\textbf n}$, there are two
first-order scalars that can be obtained from the deformation tensor:
its trace, $H^{-1}\partial v_i/\partial x_i = f(\Omega_m)\delta$, where
$f(\Omega_m)= d\log D(a)/d\log a$ is the logarithmic derivative of the
growth factor $D(a)$, and the peculiar velocity gradient along the line
of sight, $\eta=n_i n_j (\partial v_i/\partial x_j)/H$. Therefore, the
fluctuation in the \lya transmission must be given to first order by a
linear combination of $\delta$ and $\eta$, with two bias factors,
$b_\delta=\partial \delta_\alpha/\partial \delta$ and
$b_\eta=\partial \delta_\alpha/\partial \eta$, with numerical values
that depend on redshift and on the small-scale physics of the
intergalactic gas (\citealt{McDonald2000}, \citealt{McDonald2003},
\citealt{Ho09}, \citealt{Slosar2011}). Whereas galaxy surveys require
only one bias factor
to relate galaxy density to mass density fluctuations, the \lya forest
requires two of them because of the non-linear transformation from the
\lya optical depth to the observed transmission fraction, which alters
the dependence on the direction vector of the observation ${\textbf n}$.

~\par This dependence on $\delta$ and $\eta$ is only valid, however, if one
assumes that no other independent physical quantities that are
correlated on large scales can affect the value of $\delta_\alpha$;
in particular, a homogeneous ionising background intensity is assumed.
The quantity that matters for determining the \lya transmission is the
photoionisation rate, $\Gamma({\textbf x})$, obtained from the
integration over frequency of the background intensity times the cross
section. Its fluctuation is $\delta_\Gamma({\textbf x})= 
\Gamma({\textbf x})/\bar\Gamma - 1$. Including these large-scale
variations of the photoionisation rate, the total \lya transmission
fluctuation smoothed over a large scale is
\begin{equation}\label{eq:dagen}
\delta_{\alpha}(\textbf{x})=b_{\delta}\delta(\textbf{x}) +
b_{\eta}\eta(\textbf{x})+b_{\Gamma}\delta_{\Gamma}(\textbf{x}) ~.
\end{equation}
where $b_\Gamma$ is now a third bias factor for the photoionisation
rate. Therefore, the total \lya correlation depends now not only on
the correlations of $\delta$ and $\eta$ (which are related to the
primordial linear power spectrum with redshift distortions), but also
on the correlation of $\delta_\Gamma$ with itself, $\delta$ and $\eta$.
We now compute these correlations, and we will do this taking into
account two different effects: the fact that sources are clustered
and trace the mass fluctuations, and the shot noise due to the random
distribution of discrete sources.

\subsection{Source Clustering}\label{sec:sc}

~\par We assume that the sources of the ionising background have a
spatial distribution tracing the mass density field, with a bias factor
$b_s$, so the mean large-scale overdensity of sources is $\delta_s =
b_s\delta$. In addition, the ionising radiation is being absorbed by a
population of absorbers, which are Lyman limit systems as well as
absorption systems with Lyman continuum optical depths below unity that
have a comparable contribution to the overall absorption. This
population of absorbers has a large-scale distribution that is affected
by both the underlying mass density fluctuations and the radiation
intensity fluctuations. So, the absorber density fluctuation can be
written as $\delta_a = b_a\delta + b_a'\delta_{\Gamma}$. We expect
these absorbers to increase in high density regions and decrease in
response to an increased ionising intensity, so $b_a$ should be
positive and $b_a'$ should be negative.

~\par Even though the opacity to ionising photons depends on frequency,
and a detailed treatment has to include the intensity spectrum and the
combined effect of absorption and redshift modifying the background
spectrum compared to that emitted by the sources, here we shall treat
the opacity as a single quantity, neglecting the effect of redshift.
The opacity due to absorbers with density fluctuation $\delta_a$ is
$\kappa(\textbf{x}) = \kappa_0[1+\delta_a(\textbf{x})]$, where the
average mean free path for an ionising photon is $\lambda_0 =
\kappa_0^{-1}$. The radiation intensity fluctuation at a point
\textbf{x} due to the combination of all sources at any position
\textbf{x}+\textbf{r} is
\begin{equation}\label{eq:dg}
\delta_{\Gamma}(\textbf{x}) = \int d^3r\, \kappa_0
\frac{[1+\delta_s(\textbf{x}+\textbf{r})]e^{-\tau(\textbf{x},\textbf{r})}
 - e^{-\kappa_0 r}}{4\pi r^2}
\end{equation} 
where $r$ is the modulus of the vector \textbf{r}. Defining also
$\textbf{u}_r$ to be the unit vector in the direction \textbf{r}, the
optical depth from \textbf{x} to \textbf{x}+\textbf{r} is: 
\begin{align}\label{eq:tau}
\tau(\textbf{x},\textbf{r})&=  \int^r_0 dy\, \kappa_0
[1+\delta_a(\textbf{x}+y\textbf{u}_r)]\\
				&=\kappa_0 r
 \left[1+\int^r_0\frac{dy}{r}\delta_a(\textbf{x}+y\textbf{u}_r)\right] ~.
 \nonumber
\end{align} 
Neglecting second order terms in $\delta_s$ and $\delta_a$,
equation (\ref{eq:dg}) is simplified to
\begin{align}
\delta_{\Gamma}(\textbf{x})& =\int \frac{d\textbf{u}_r}{4\pi}
\int^{\infty}_0dr \kappa_0\, e^{-\kappa_0 r} \\
						& \cdot
\left[\delta_s(\textbf{x}+r\textbf{u}_r) - 
\kappa_0 \int^r_0 dy \, \delta_a(\textbf{x}+y\textbf{u}_r)\right] ~. \nonumber
\end{align} 

~\par For the second term involving the absorbers, the order of the
integrals over $r$ and $y$ can be inverted, and we find: 
\begin{align}
\int^{\infty}_0 & dr \kappa_0^2 e^{-\kappa_0 r}
\int^{r}_0dy\delta_a(\textbf{x}+y\textbf{u}_r) \\
&=\int^{\infty}_0  dy\, \kappa_0^2\delta_a(\textbf{x}+y\textbf{u}_r)
\int^{\infty}_y dr e^{-\kappa_0 r}\nonumber \\
&=\int^{\infty}_0 dy\, \kappa_0 e^{-\kappa_0 y}
\delta_a(\textbf{x}+y\textbf{u}_r) ~, \nonumber
\end{align}
and so finally, changing the name of the dummy variable $y$ back to $r$,
and reexpressing the integral in terms of the variable \textbf{x'=x+r},
\begin{equation}
\label{eq:dg1}
\delta_{\Gamma}(\textbf{x})=\int \frac{d \textbf{x'}}{4\pi r^2}
[\delta_s(\textbf{x'})-\delta_a((\textbf{x'})]\kappa_0 e^{-\kappa_0 r} ~.
\end{equation}
This result is easy to understand, because an absorber actually acts
in the same way as a negative source in this linear regime. 

~\par We now replace $\delta_{\Gamma}(\textbf{x})$ and
$\delta_s(\textbf{x'})$, $\delta_a(\textbf{x'})$ in equation
(\ref{eq:dg1}) by their Fourier transforms, invert the order of the
integrals over \textbf{k} and \textbf{x'}, and do the integral over
\textbf{x'}, to find that the Fourier transforms are related by
\begin{align}\label{eq:dgb}
\delta_{\Gamma}(\textbf{k}) &= [\delta_s(\textbf{k})-\delta_a(\textbf{k})]\,
 W{\left(\frac{k}{\kappa_0}\right)}\\ 
&= [(b_s-b_a)\delta(\textbf{k}) - b'_a\delta_{\Gamma}(\textbf{k})]\,
 W{\left(\frac{k}{\kappa_0}\right)} ~, \nonumber
\end{align}
where 
\begin{equation}\label{eq:kerd}
W(s)=\int^{\infty}_0 dx\, \frac{e^{-x}\sin{(sx)}}{sx} =
 \frac{{\rm arctan}(s)}{s} ~.
\end{equation}

~\par Writing now equation (\ref{eq:dagen}) in Fourier space, and suppressing
dependences on $k$ for brevity, the correlation of the Fourier modes of
the \lya transmission fluctuation is
\begin{align}
\langle\delta_{\alpha}\delta_{\alpha}\rangle & =
 b^2_{\delta}\langle\delta\delta\rangle + b^2_{\eta}\langle\eta\eta\rangle
+ b^2_{\Gamma}\langle\delta_{\Gamma}\delta_{\Gamma}\rangle \\
	& + 2b_{\delta} b_{\eta}\langle\delta\eta\rangle
+ 2b_{\delta}b_{\Gamma}\langle\delta\delta_{\Gamma}\rangle + 
 2b_{\eta}b_{\Gamma}\langle\eta\delta_{\Gamma}\rangle ~. \nonumber
\end{align}
Using the linear redshift distortion theory of \cite{Kaiser1987}, the
Fourier modes of $\delta$ and $\eta$ are related by
$\eta=f(\Omega_m)\mu_k^2\delta$, where
$\mu_k\equiv {\textbf n}\cdot {\textbf k}/k$, and the power spectrum
without including the radiation term $\delta_\Gamma$ can be written as
usual in the form $b_\delta^2(1+\beta\mu_k^2)^2$, where the redshift
distortion parameter is $\beta=f(\Omega_m)b_\eta/b_\delta$. When the
radiation term is included and expressed as a function of $\delta$
using equation (\ref{eq:dgb}), we find that the total \lya power
spectrum is given by
\begin{equation}\label{eq:pwsp}
P_{\alpha}(k,\mu_k) = P_L(k)\, b_{\delta}^{'2}(k)
\left[1+\beta'(k)\mu_k^2\right]^2 ~,
\end{equation}
where
\begin{equation}\label{eq:bias}
b_{\delta}'(k)= b_{\delta} + b_{\Gamma}
\frac{(b_s-b_a)W(k/\kappa_0)}
 {1+ b'_aW(k/\kappa_0)} ~,
\end{equation}
and $\beta'(k)= b_\delta \beta/b_\delta'(k)=
b_\eta f(\Omega_m)/b_\delta'(k)$.

~\par Therefore, the effect of the photoionisation rate fluctuations that is
induced by the clustering of sources and absorbers is to modify the bias
factor and redshift distortion parameter in the power spectrum,
replacing them with the
effective values $b_\delta'$ and $\beta'$ that are scale dependent,
while their product $b_\delta'\beta'=b_\delta\beta$ remains fixed. At
small scales, $W$ is very small and the bias factor has its usual value
$b'_{\delta}=b_{\delta}$. But in the limit of large scales, $W$
approaches unity and $b_{\delta}'$ reaches the asymptotic value of
$b_{\delta}+b_{\Gamma}(b_s-b_a)/(1+b_a')$.

~\par We now interpret physically the variation of the effective bias
$b_\delta'$ with the Fourier scale $k$. We mention first that any
realistic model for the absorbers needs to have $0 > b_a' > -1$: the
density of absorbers (which we identify with the observed population
of Lyman limit systems, as well as lower column density systems that
also contribute to the global absorption of ionising photons) needs to
decrease with $\delta_\Gamma$ as the
increased photoionisation reduces the size of the absorbing regions,
but the relative fluctuation in absorbers cannot be reduced faster than
that in the ionising intensity because this would imply a runaway
unstable process where any slight increase in emission leads to an
arbitrarily large increase in the mean free path and the ionising
intensity as the absorbers are completely ionised. Moreover,
the sign of $b_{\delta}$ is negative while that of $b_{\Gamma}$ is
positive, so if $b_s>b_a$, the effective bias factor $b_{\delta}'$
decreases in absolute value with scale. The simple interpretation is
that on scales large compared to the mean free path of ionising
photons, denser regions also have a greater ionising intensity, and so
the corresponding increase of \lya absorption that is caused by the
higher density is reduced. If
$b_\Gamma(b_s-b_a)/(1+b_a')$ is larger than $-b_\delta$, then the
value of $b_\delta'$ is actually positive in the limit of large scales.
Ignoring for now the effect of the peculiar velocity gradient (we return
to this in section 3), this
means that the effect of the higher ionising intensity overwhelms that of
the higher mass density, causing denser regions to have an increased
\lya transmission (or reduced absorption), opposite to the behavior
on small scales. In this case, there needs to be a critical scale $k_r$
where $b_\delta'$ has a root, and the only surviving term for
the power spectrum in equation (\ref{eq:pwsp}) is $P_\alpha(k_r)=
P_L(k_r) (b_\delta \beta \mu_k^2)^2$.
In practice, the power spectrum near $\mu_k=0$ can never quite go down
to zero owing to the shot noise from individual sources of radiation, as
discussed below, as well as non-linear effects from small scales which
we are not including here, but a change of sign of $b_\delta'$ as a
function of scale still implies the presence of a dip in the power
spectrum at small $\mu_k$ which should be measurable in the
observations.

\subsection{Shot noise from individual sources}\label{sec:is}

~\par Our treatment so far includes only the correlation of sources and
absorbers with the matter density fluctuations. Next, we consider the
term that is added to the correlation function because of the
shot noise from individual sources. We assume that all the ionising
sources emit their radiation isotropically at a constant luminosity.
As we shall discuss below, this assumption is crucial for our
computation of the shot noise term, even though it does not affect
the source clustering term calculated above. We start defining the
source luminosity function, per unit of volume, as $\Phi(L)$. The mean
emissivity of the ionising sources is
\begin{equation}
\epsilon_q = \int_0^\infty dL\Phi(L)L ~.
\end{equation}
If all the sources have the same luminosity, the power spectrum of the
relative emissivity fluctuations is simply equal to the inverse of the
source density, $1/n_s= L/\epsilon_q$. With a distribution of
luminosities, the fraction of the emissivity provided by sources of
luminosity $L$ is $L\Phi(L)\, dL/\epsilon_q$, and therefore the overall
power spectrum is equal to the constant
\begin{equation} \label{eq:cdef}
4\pi C=\int_0^\infty dL\, \frac{\Phi(L)L^2}{\epsilon_q^2} ~,
\end{equation}
where we have introduced the factor $4\pi$ in the definition of $C$
for later convenience.
The intensity correlation can now be obtained by multiplying
by the kernel in equation (\ref{eq:kerd}), and applying the Fourier
transform. It is also instructive, however, to directly compute the
correlation function by considering the correlated intensity at two
spatial positions separated by a distance $x$ caused by the flux that
arrives at the two points from the same individual sources.

~\par
\begin{figure}
\includegraphics[width=1\linewidth]{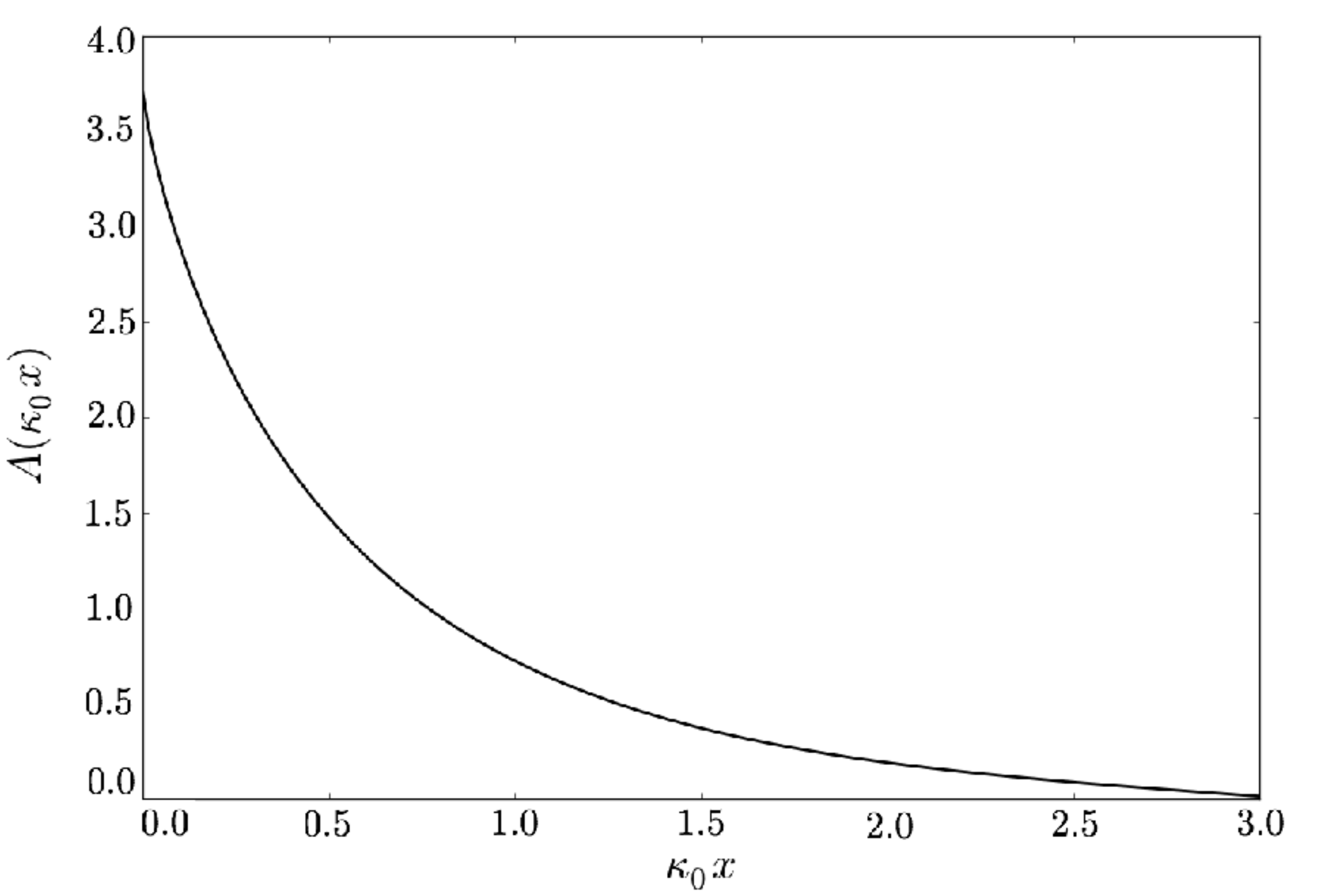}
 \caption{The dimensionless function $A$ defined in equation (\ref{eq:dless}),
giving the shape of the intensity correlation function due to a set of randomly
distributed, isotropic and constant sources of radiation.}
 \label{fig:dless}
\end{figure}

 We choose one of the two spatial positions to be at the origin of
coordinates, ${\bf r}=0$, and the other one to lie on the x-axis at a
distance $x$. The ionising intensity fluctuation at the origin is
\begin{equation}
\delta_{\Gamma}(\mathbf{r}=0)=\sum_i\frac{L_i}{4\pi r^2_i}
\frac{\kappa_0e^{-\kappa_0 r_i}}{\epsilon_q}-1 ~,
\end{equation}
where the sum is over each source $i$ located at a distance $r_i$ from
the origin. The intensity fluctuation at ${\bf x}$ is similarly
expressed, replacing $r_i$ by $| {\bf r}_i - {\bf x} |$. Using the
fact that the probability per unit of volume to find a source of
luminosity $L$ within $dL$ at any point ${\bf r}_i$ is $\Phi(L)\, dL$,
the correlation function of $\delta_\Gamma$ is then obtained as
\begin{align}\label{eq:IS}
\xi_{\Gamma}(x) = &
\langle\delta_{\Gamma}(\mathbf{r}=0)\delta_{\Gamma}(\mathbf{x})\rangle= \\
 & {C\over 4\pi} \int d^3r
\frac{\kappa_0^2 e^{-\kappa_0(r+|\mathbf{r}-\mathbf{x}|)}}
{r^2|\mathbf{r}-\mathbf{x}|^2} ~. \nonumber
\end{align}
We compute this integral by transforming $\bf r$ to spherical
coordinates. Defining $\mu= {\bf r\cdot u}_x/r$, where ${\bf u}_x$
is the unit vector along the x-axis, and changing $r$ to the variable
$s=r/x$, the result is
\begin{align}\label{eq:2dIS}
\xi_{\Gamma}(x) =& C \kappa_0^2 \int^\infty_0 \frac{ds}{x} \cdot \\
 & \int^1_{-1} d\mu \,
 { \exp \left[-\kappa_0 x \left( s+\sqrt{1+s^2-2s\mu} \right)\right] \over
    1+s^2-2s\mu} ~. \nonumber
\end{align}
This can be further reexpressed in terms of the exponential integral
function, $E_i(x)=-\int_{-x}^{\infty} dt\, e^{-t}/t$. The final
expression for the intensity correlation function is
\begin{equation} \label{eq:xishot}
\xi_{\Gamma}(x)= \frac{C\kappa_0^2}{x}A(\kappa_0 x) ~,
\end{equation}
where the dimensionless function $A$ is (replacing $\kappa_0 x=\tau$
and $\kappa_0 r= \rho$)
\begin{align}\label{eq:dless}
&A(\tau)= \int_0^{\infty}d\rho\, \frac{e^{-\rho}}{\rho} \cdot \\
& \left[ E_i\left(- \rho \sqrt{1+\frac{\tau^2}{\rho^2}}\right) -
 E_i\left(- \rho \left|1-\frac{\tau}{\rho}\right|\right)\right] ~. \nonumber
\end{align}
The function $A$ is plotted in Figure \ref{fig:dless}.

~\par The sources that dominate the fluctuations in the ionising radiation
intensity are the most luminous ones, which are well known from
observations of the quasar luminosity function. We now estimate the
constant $C$ from recent measurements of the quasar luminosity function
by \citep{Ross2013}, who used the BOSS survey of the SDSS-III
Data Release 9 \citep[see][]{Eisenstein2011,Ahn2012,Dawson2013}. The
quasar luminosity function was fitted to a double power-law of the form,
\begin{equation}\label{eq:qlf}
 \Phi_q(L)dL=\frac{\Phi_{*}/L_*}{(L/L_*)^{-\alpha} +
(L/L_*)^{-\gamma}}\, dL ~.
\end{equation}
The values of the fitted parameters obtained by \citet{Ross2013} are
$\alpha=1.52$, $\gamma=3.10$, and
$\Phi^{(R)}_{*}=10^{-6.37}\mpc^{-3}{\rm mag}^{-1}$, where they defined
$\Phi^{(R)}_{*}$ to be a number density of quasars per unit of absolute
magnitude. We convert their value
$\Phi^{(R)}_{*}$ to our cosmological model, at $z=2.25$ (the model used
by \citealt{Ross2013} had $\Omega_m=0.30$ and $H_0=70 \kms\mpc^{-1}$)
and to our units, finding $\Phi_{*}= 1.42\times 10^{-6} (\hmpc)^{-3}$.
With these
numbers, we compute the quantity $C_q$ for quasars using equation
(\ref{eq:cdef}), and we find $C_q=9.5\times 10^4 (\hmpc)^3$. Note
that the value of $C_q$ is independent of $L_*$, and depends only
on $\Phi_*$ and the shape of the luminosity function. The value of
$C_q$ diverges as $\gamma$ approaches 3, so the fitted value of
$\gamma=3.1$ from \cite{Ross2013} implies a large uncertainty of $C_q$
depending on the exact shape of the luminosity function at the
high luminosity end.

~\par The intensity of the ionising background is likely to have a
contribution from galaxies, in addition to quasars. Assuming that
all galaxies are much less luminous than the quasars that contribute
appreciably to the constant $C_q$, then the faint galactic sources
can increase the mean intensity, but can be neglected for the
fluctuations, i.e., their contribution to the integral
$\int dL\, \Phi(L) L^2$ is negligible. In that case, equation
(\ref{eq:cdef}) implies that the emissivity power spectrum
amplitude is determined by the constant
\begin{equation}\label{eq:norm}
C=\frac{\epsilon_q^2 C_q}{(\epsilon_g + \epsilon_q)^2}  ~,
\end{equation}
where $\epsilon_g$ and $\epsilon_q$ are the emissivities of galaxies
and quasars. The correlation due to shot noise is therefore reduced
as the contribution from galaxies to the ionising intensity is
increased.

 ~\par Apart from the effect of galaxies, the amplitude of the intensity
correlation function contributed by shot noise from the observed
quasars that is obtained from equation (\ref{eq:xishot}) ought to be
considered as an upper limit only. The reason is that real quasars
are likely to emit anisotropically and to be highly variable in their
luminosity on the light-crossing time of the cosmological scales at
which the correlations are being measured. The effects of the variability
and anisotropy of the emission from individual quasars can be highly
complex and difficult to model, but an order-of-magnitude estimate
can be made by assuming the opposite limit in which quasars emit in
very narrow cones and short bursts. If $f_\Omega$ is the fraction of
the solid angle over which the light of a quasar is emitted, and $f_t$
is the fraction of time during which a quasar is shining, and we
assume that the quasar luminosity is zero
outside of this fraction of solid angle and time, then the correlation
function $\xi_\Gamma$ in equation (\ref{eq:xishot}) is reduced by a
factor $f_\Omega f_t$, because assuming that a source is being
observed at a point ${\bf r}=0$, the probability that it is also
observed at a point ${\bf x}$ with the same luminosity is only
$f_\Omega f_t$.

~\par The total correlation function of the \lya transmission is equal to
the sum of the term due to source clustering, from equation
(\ref{eq:pwsp}), and the shot noise term in equation (\ref{eq:xishot})
multiplied by $b_\Gamma^2$.

\subsection{Values of the bias parameters}

~\par We now discuss how the values of the various bias parameters that
have appeared in our derivation of the radiation effects in the \lya
correlation function can be estimated. All the bias factors should
generally depend on redshift, and our discussion here will be focused
at $z=2.25$, the redshift near which the observations of \lya
correlations have so far been done with the BOSS survey.

~\par We start with the bias factor that relates the ionising intensity
fluctuations to the \lya transmission fluctuations. As described in
\cite{Font13} (see their equation 5.4), this bias factor can
easily be calculated from its definition as $b_\Gamma =
\partial\delta_\alpha / \partial \delta\Gamma$, where the values
$\delta_\alpha$ and $\delta_\Gamma$ are smoothed over sufficiently
large, linear scales, if the true, unsmoothed distribution of the \lya
transmission is known, and under two additional assumptions:
photoionisation equilibrium in a highly ionized medium (neglecting any
contribution from collisional ionisation), and that any changes of
temperature and hydrodynamic evolution of the intergalactic gas with the
ionising background intensity can be neglected (in a more detailed
treatment, small temperature variations would likely be induced by
changes in the spectral shape of the ionising background rather than its
intensity). In this case, the effect of the ionising intensity
fluctuations is simply to divide the \lya optical depth by a factor
$1+\delta_\Gamma$ at every pixel in the spectrum. If $P(F)$ is the
unsmoothed probability distribution of $F$, one obtains
\begin{equation}
  b_\Gamma= - {1\over \bar F} \int_0^1 dF\, P(F)\, F\, \log(F) ~.
\end{equation}
At $z=2.25$, the distribution of the \lya optical depth, $\tau=-\log F$,
can be approximated as a log-normal function, constrained to produce a
mean transmission $\bar F=0.8$ and a dispersion in the transmission
$\sigma_F=0.124$. This distribution yields a value $b_\Gamma\simeq 0.13$.

~\par The values of $b_\delta$ and $b_\eta$ have to be measured
observationally. They are related to the redshift distortion parameter
by $\beta=f(\Omega_m) b_\eta/b_\delta$ \citep{Kaiser1987}. Measurements
from the \lya forest correlation in the scale range $10-60 \hmpc$ in
\cite{Slosar2011} resulted in a good measurement of
$b_\delta(1+\beta)=-0.336\pm 0.012$ at $z=2.25$, with a more
poorly constrained $\beta\sim 1$. We shall assume $\beta=1$, which is
also favored by the measurement of cross-correlations (see
\citealt{Font12}, \citeyear{Font13}), and $b_\delta=-0.17$ (so
$b_\eta=\beta b_\delta/f(\Omega_m)\simeq -0.17$; note that the
negative sign of $b_\delta$ and $b_\eta$ results from the convention
that $\delta_\alpha$ is a transmission fluctuation, which is therefore
negative when the mass density perturbation is positive).
These observational results may change in the future since they were
obtained by neglecting the radiation effects that are examined here,
and they are subject to other possible systematic errors
\citep[e.g.,][]{FontMiralda2012}.

~\par For the bias of the sources, quasars have had their bias factor
measured from their auto-correlation
\citep[][and references therein]{White2012}
and cross-correlation with the \lya forest
\citep{Font13}, resulting in values in the range $3.5$
to 4. The actual bias of the sources, however, depends also on the
contribution that galaxies make to the ionising background intensity
and on the bias factor of these galaxies. If the bias factors of quasars
and galaxies are $b_q$ and $b_g$, then $b_s=(\epsilon_q b_q +
\epsilon_g b_g)/(\epsilon_q+\epsilon_g)$. Galaxies are on average
associated with lower-mass halos than quasars, so their bias factor
should be smaller and therefore $b_s$ should be lower than $b_q$.

~\par The population of absorbers determining the mean free path of the
ionising radiation is dominated by systems that have an optical depth of
order unity at the Lyman limit, with column densities $\sim
1.6\cdot 10^{17}\text{cm}^{-2}$. Note that only systems with column
densities above this value are usually referred to as Lyman limit
systems, but absorbers of lower column density are about equally
important \citep[e.g.,][]{MO90,HM96}. The bias factor has only
been measured for systems of higher column densities, the damped \lya
systems \citep{Font12}, and a value $b_a\simeq 2$
was obtained. Lyman limit systems are of lower column density than
the damped systems but they should have a similar bias factor if halos
of all masses give rise to the same distribution of hydrogen column
densities, depending on the impact parameter. However, a population of
low-mass halos might exist in which the self-shielded gas does not reach
as high column densities as in high-mass halos, which would then reduce
the mean bias factor of the Lyman limit systems. Both $b_s$ and $b_a$
are therefore rather uncertain. The effective \lya forest bias
depends only on the difference $b_s-b_a$ (see equation \ref{eq:bias}),
for which we shall assume a fiducial value $b_s-b_a=1$.

~\par Finally, the bias factor controlling the response of the absorbers
to changes in the ionising intensity can be related to the column
density distribution of Lyman limit systems, which we model as a
power-law, $f(N_{HI}) \, dN_{HI}\propto N_{HI}^{-a}\, dN_{HI}$. This
implies a radial profile of the column density in spherical halos
$N_{HI}\propto r^{2/(1-a)}$. In
photoionisation equilibrium, the column density at a fixed radius $r$
outside the region where the gas starts to self-shield will vary in
proportion to the inverse of the photoionisation rate. The self-shielding
radius $r_s$ occurs at a fixed column density, and will therefore
change as $r_s\propto \Gamma^{(1-a)/2}$ in response to a change of the
external ionising intensity. The cross section to produce a Lyman
limit system scales as $r_s^2$, so the number of absorbers that are
intercepted per unit length should scale as $\Gamma^{1-a}$. For small
changes in $\Gamma$ this implies a bias factor $b_a'=1-a$. We shall use
here $a=5/3$ (the value corresponding to a gas density profile
$\rho_g \propto r^{-2}$, and $N_{HI}\propto r^{-3}$), and therefore
$b_a'=-2/3$.

\begin{figure*}
\includegraphics[width=1.\textwidth]{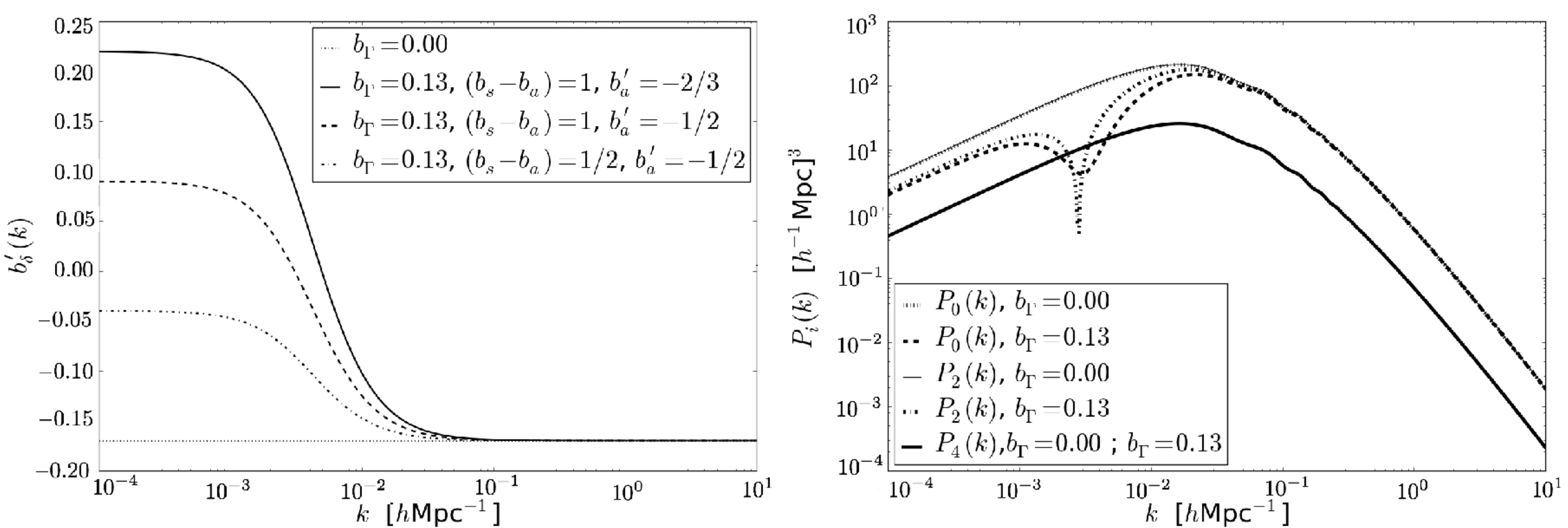}
\caption{Effective bias (left) and power spectrum (right) of the
 \lya forest, for several values of the bias
parameters regulating the clustering strength of sources minus absorbers
($b_s-b_a$) and the response of the absorbers to the ionising intensity
($b_a'$). The dotted line is for no radiation effects, and the solid
line in the left panel is for our fiducial radiation model.
The mean free path is fixed to $\lambda_0=300 \hmpc$. The
right panel shows the monopole, quadrupole and hexadecapole of the
power spectrum, for the cases of no radiation fluctuations and for
our fiducial radiation model.}\label{pfig}
\end{figure*}

~\par With the values we adopt here for the fiducial model, the value
of $b_\delta'(k)$ in equation (\ref{eq:bias}) is plotted in the left
panel of Figure \ref{pfig}, as the solid line.
Here and in the rest of the paper, we use
a mean free path $\lambda_0 = 300 \hmpc$ as our fiducial value.
Observational estimates of the mean free path of an ionising
photon for being absorbed by hydrogen in the intergalactic medium or
Lyman limit systems give a value $\lambda_0 \simeq 350 \hmpc$ \citep{Rudie13}.
Our approximate treatment in section 2.2 neglects the redshift of the photons,
which effectively acts in the same way as an additional source of
opacity with a comoving mean free path of the order of the horizon,
$c(1+z)/H(z)\simeq 3000 \hmpc$. The effective overall mean free path
is therefore close to our fiducial value of $300 \hmpc$.

 ~\par The two other curves in Figure \ref{pfig} are for variations of
the bias values that will be used in section 3. For our fiducial model,
$b_\delta'(k)$ is positive in the limit of large scales and negative
at small scales, and therefore changes sign at a critical scale $k_r$.
Depending on the uncertain values of all the bias factors we have
discussed, this critical scale can have very different values, and is
likely to vary substantially with redshift.

\subsection{Fluctuations due to helium reionisation}

~\par  In addition to the intensity of the ionising background, the
intergalactic medium may be affected by other physical elements that are
correlated over large scales. Here we consider as another possibility
the imprint that may have been left by helium reionisation in the
temperature of the intergalactic gas. At the mean baryonic density of
the universe, the recombination time at $z=2.25$ is much longer than the
age of the universe, and so is therefore the cooling time of
photoionised gas. As helium is doubly ionised for the first time,
probably by luminous quasars at $z\simeq 3$
\citep[][and references therein]{Worseck2011}, the gas is heated to a
spatially variable temperature depending on the spectrum and luminosity
of the sources producing the ionisation fronts that eventually overlap
when reionisation ends \citep[e.g.,][]{MR1994,McQuinn2009}. The long
cooling time then implies that the gas temperature at every spatial
location may keep a memory of the time at which helium reionisation
occurred, or the spectral shape of the sources, or other characteristics
that were imprinted at the reionisation time. If the temperature
fluctuates according to $\delta_T = b_T \delta_{\Gamma e}$, where
$\delta_{\Gamma e}$ is an intensity fluctuation of the HeII-ionising
radiation that was present at
the reionisation time arising from sources that may long have been
dead, then the observed \lya transmission in hydrogen would vary
as $\delta_\alpha = b_e \delta_{\Gamma e}$ owing to the dependence
of the recombination coefficient on temperature, which follows the
approximate relation $\alpha_{rec}(T)\propto T^{-0.7}$. Using similar
arguments as in section 2.3 for deriving $b_{\Gamma}$, we can infer
that $b_e = 0.7 b_\Gamma b_T$.

 ~\par If the helium-ionising radiation intensity follows the same behavior
as the radiation that ionised hydrogen, with a different mean
opacity $\kappa_{0e}$, a similar derivation as in section 2.1 shows
that the total power spectrum that includes also the clustering term
for the sources inducing the helium-reionisation perturbations is the
same as in equation (\ref{eq:pwsp}), with the new bias factor
\begin{align}\label{eq:biashe}
b_{\delta}'(k) =& b_{\delta} + b_{\Gamma}\frac{(b_s-b_a)W(k/\kappa_0)}{1+ b'_aW(k/\kappa_0)} \nonumber \\
 +& b_e \frac{(b_{se}-b_{ae})W(k/\kappa_{0e})}{1+ b'_{ae}W(k/\kappa_{0e})} ~,
\end{align}
where the bias factors with the additional subscript e are the
analogous ones for helium to those that were described for hydrogen
in section 2.1.

\section{Results}

 ~\par The correlation function of $\delta_\alpha$ is obtained from the
Fourier transform of the power spectrum in equation (\ref{eq:pwsp}),
which includes the source clustering term. To include the shot noise
term, the correlation $\xi_\Gamma$ from equation (\ref{eq:xishot})
multiplied by $b_\Gamma^2$ must be added. Following the formalism and
notation of \citet{Kirkby13} (see their section 2.2), the multipole
terms of the power spectrum of equation (\ref{eq:pwsp}) have their usual
form with the new scale-dependent bias factor $b_\delta'(k)$ and
redshift distortion parameter $\beta'(k)$
\citep{Kaiser1987,Hamilton1992}: 
\begin{equation}\label{eq:mutiPS2}
       P_{\ell,\alpha}(k)=P_L(k)\, b_{\delta}^{\prime2}(k)\, C_{\ell}[\beta'(k)] ~,
\end{equation}
where
\begin{align}
 C_0 =& 1+{2\over 3} \beta'(k) + {1\over 5} \beta'^2(k) ~, \nonumber \\
 C_2 =& {4\over 3} \beta'(k) + {4\over 7} \beta'^2(k) ~, \nonumber  \\
 C_4 =& {8\over 35} \beta'^2(k) ~. \\
\end{align}

~\par  The multipoles of the real space \lya correlation function are
\begin{equation}\label{eq:multiCF}
\xi_{\ell} = \frac{i^{\ell}}{2\pi^2}\int_0^{\infty} dk\,
 k^2j_{\ell}(kr)\, P_{\ell,\alpha}(k) ~,
\end{equation}
where $j_{\ell}$ are the spherical Bessel functions. Note that these
correlation multipoles are
no longer given by the equations in \citet{Hamilton1992} because of the
scale dependence of $b_\delta'$ and $\beta'$, except for the
hexadecapole which does not change because
$b_\delta' \beta' = b_\delta \beta$. We have used a Fast Fourier
Transform method to calculate these multipoles numerically.

 ~\par As shown in the left panel of Figure \ref{pfig}, the bias factor
$b_\delta'(k)$ in our fiducial radiation model changes sign at a
critical scale $k_r\simeq 0.005\, h/{\rm Mpc}$. The monopole,
quadrupole and hexadecapole of the power spectrum are well defined and
non-zero at $k=k_r$ because $b_\delta' \beta'$ is constant.
At $k < k_r$, the redshift distortion parameter $\beta'(k)$ is
negative. While the monopole is always positive for any value of
$\beta'$, the quadrupole is zero when $\beta'(k)=-7/3$, and becomes
negative at small $k$, when $-7/3 < \beta' < 0$. This is seen in the
right panel of Figure \ref{pfig}, where the monopole, quadrupole and
hexadecapole of the power spectrum are shown for the no radiation
case, and the case that includes the source clustering effect for
our fiducial values of the radiation bias parameters (the power
spectrum is computed for the Cold Dark Matter model with the parameters
mentioned in the introduction; all results for power spectra and
correlation functions in this paper are shown at $z=2.25$). The monopole
has a dip near $k=k_r$, and the quadrupole has a root at the slightly
smaller value of $k$ where $\beta'=-7/3$. The reason for this behaviour
is that when $b_\delta'$ is negative on large scales, high-density regions
produce reduced absorption owing to the larger ionising intensity
that overwhelms the gas density effect, but the gradient of peculiar
velocity counteracts that, resulting in a negative quadrupole for the
power spectrum.

\begin{figure*}
\includegraphics[width=1.\textwidth]{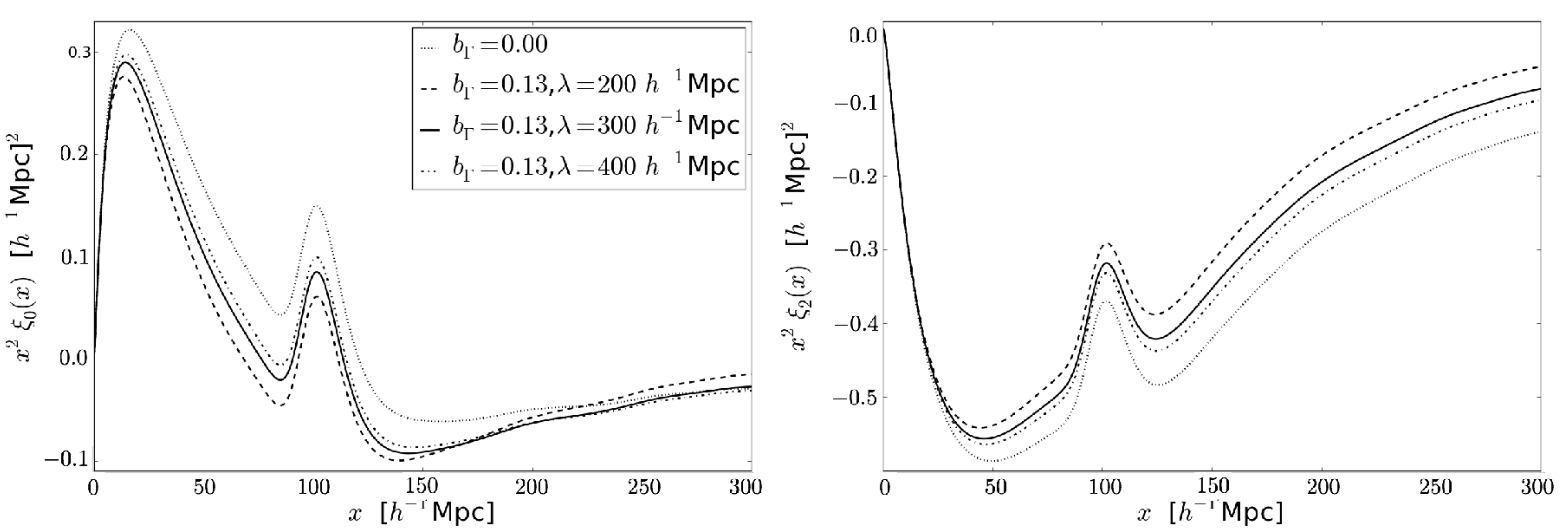}
\caption{Monopole ($\ell=0$, left) and quadrupole ($\ell=2$, right) of
the \lya autocorrelation function. The dotted line is with no radiation
effects, and the other three lines include them with our fiducial
value of the bias factors in section 2.3 and three different values of
the mean free path.}\label{fig2}
\end{figure*}

~\par  The monopole and quadrupole terms in the \lya correlation for our
fiducial case of the values of the bias parameters discussed in section
2.3 are shown in the two panels of Figure \ref{fig2}, multiplied for
convenience by $x^2$. The dotted line is for a uniform ionising
background. The
well known Baryon Acoustic Oscillation peak appears at its
characteristic scale of $\sim 100 \hmpc$. The
dashed, solid and dash-dot lines include the source clustering effect,
with no shot noise,
for three different values of the comoving mean free path: 200, 300 and
$400 \hmpc$ (our fiducial value used in all other figures is
$\lambda_0=300 \hmpc$). This mean
free path decreases rapidly with redshift, which should change the way
that intensity fluctuations modify the correlation function as the
redshift increases.

~\par  The radiation from clustered sources adds a broadband term
that is negative in the monopole and positive in the quadrupole. This
is because the absolute value of the bias $b_\delta'$ in equation
(\ref{eq:bias}) is reduced at scales small compared to $\lambda_0$, as
long as $b_s > b_a$. The shorter the mean free path, the larger the
radiation effects. On scales larger than the mean free path, the
impact of the radiation on the monopole becomes positive. The effects
are predicted to be relatively large, and they should be measurable as
long as the broadband shape can be retrieved from the data without
substantial systematic errors caused by the quasar continuum fitting
operation. In the
observations reported so far, broadband terms were marginalised over
\citep{Busca2013,Slosar2013,Font14} and therefore the effect of the ionising
intensity fluctuations would not have been detected. Note that the
position of the BAO peak is practically not affected; even if the peak
shifts by a small amount owing to the addition of the radiation
effects, any such shift should be further reduced when fitting with a
parameterised broadband term.

\begin{figure*}
\includegraphics[width=1.\textwidth]{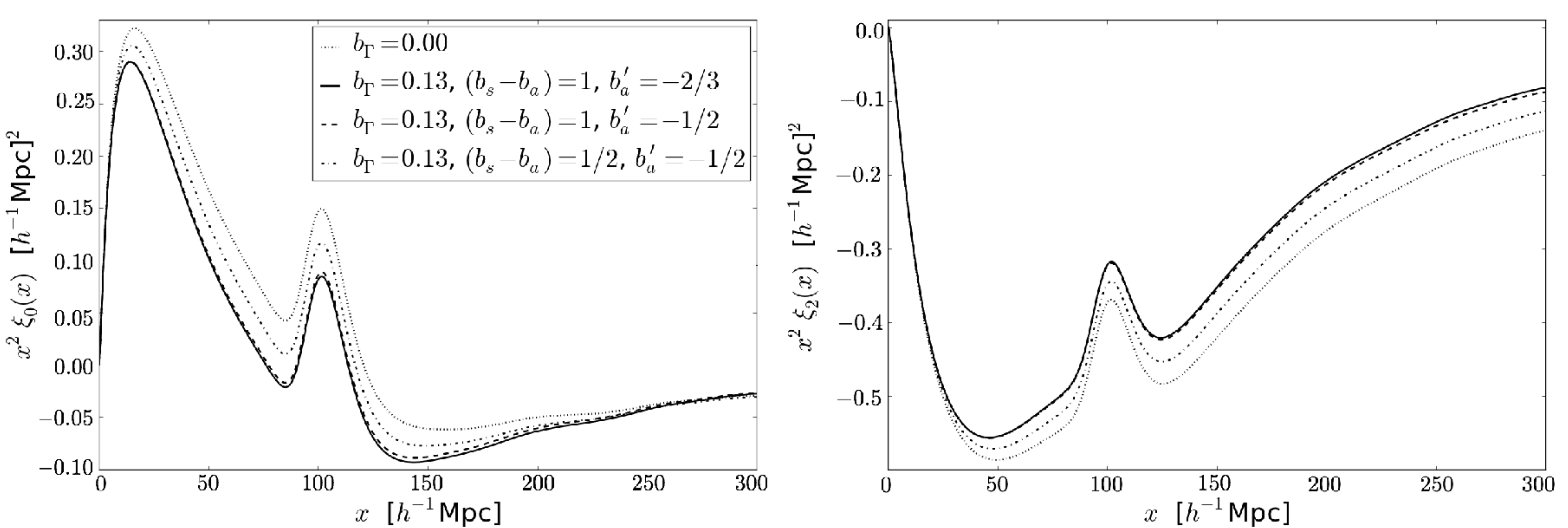}
\caption{Monopole (left) and quadrupole (right) of
the \lya autocorrelation function, for different values of the bias
parameters regulating the clustering strength of sources minus absorbers
($b_s-b_a$) and the response of the absorbers to the ionising intensity
($b_a'$). The dotted line is with no radiation
effects. The mean free is fixed to $\lambda_0=300 \hmpc$. }\label{fig3}
\end{figure*}

 ~\par Figure \ref{fig3} shows how the radiation effects vary with some
of the bias parameters (again, with the monopole on the left panel and
the quadrupole on the right panel). The dotted line (no radiation
effects) and the solid line (for the fiducial values of the bias
parameters in section 2.3) are the same as in Figure \ref{fig2}. The
dashed line shows that the radiation effect is very insensitive to
$b_a'$ on scales small compared to $\lambda_0$. The
dash-dot line has a reduced value of $b_s-b_a$, and shows that the
radiation effect is basically proportional to this bias difference
between sources and absorbers.

\begin{figure}
\includegraphics[width=1\linewidth]{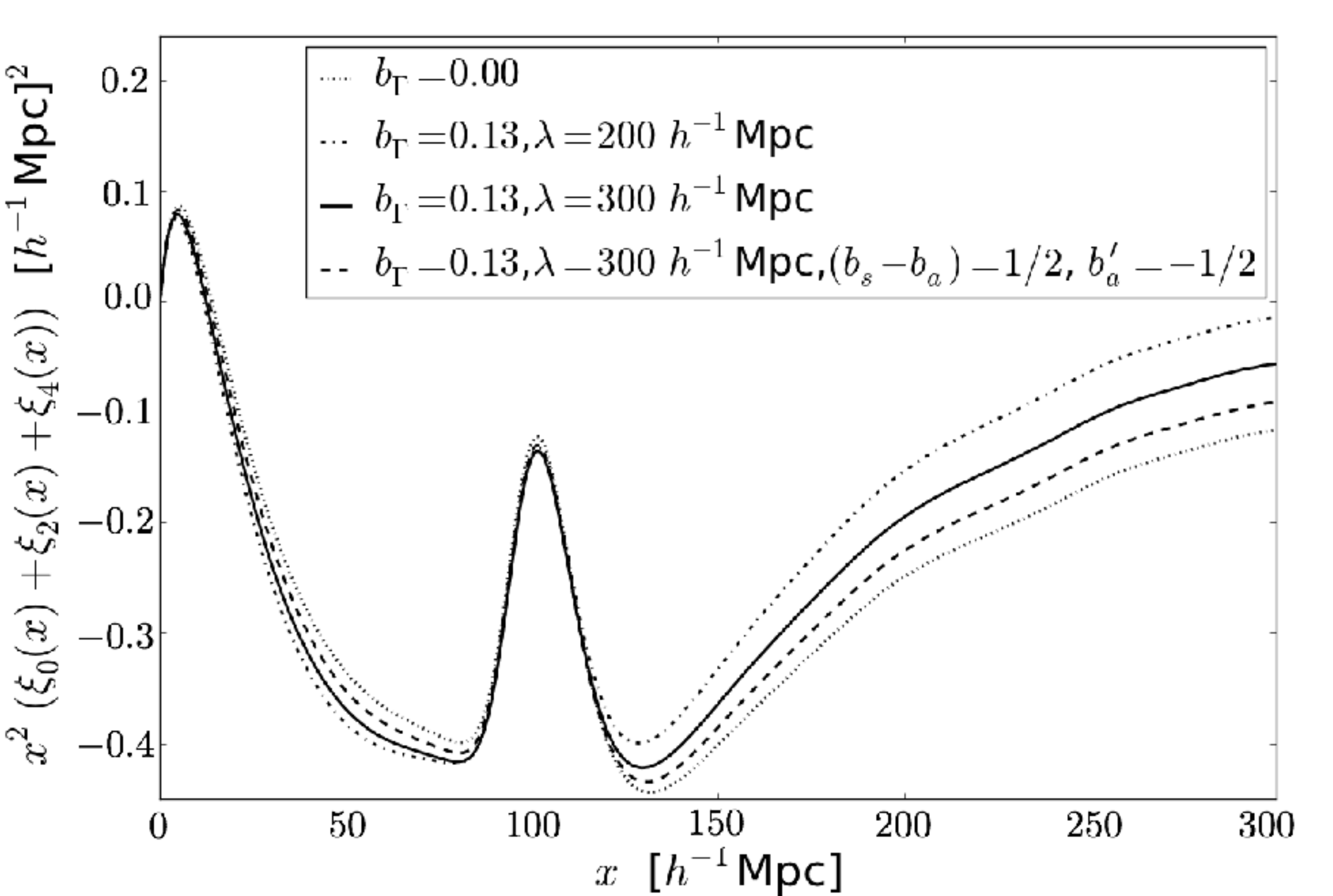}
\caption{The sum of the monopole, quadrupole and hexadecapole of the \lya
autocorrelation function. The dotted and solid lines are as in Figures
2 and 3. The dash-dot line changes the mean free path to $200 \hmpc$,
and the dashed line shows the effect of changing the bias parameters
from our fiducial model to the values indicated in the legend.}
\label{fig:monoquad}
\end{figure}

%%%%%%%%%%%%%%%%%%%%%%%%%%%%%%%%%%%%%
%%%%%%%Here change comment on the hexadecapole%%%%%%%%
%%%%%%%%%%%%%%%%%%%%%%%%%%%%%%%%%%%%%

~\par  Observations of the correlation function (or one-dimensional power
spectrum) can also be done exclusively on the line of sight
\citep{McDonald2006,Palanque2013}. The correlation function along the
line of sight is equal to the sum of all the multipoles. 
The result is shown in Figure \ref{fig:monoquad} for some of the same
models shown in
Figures \ref{fig2} and \ref{fig3}. This figure shows that the
correlation along the line of sight on scales small compared to the
mean free path is much less affected by the radiation fluctuations than
the three-dimensional correlation.

~\par
\begin{figure}
\includegraphics[width=1\linewidth]{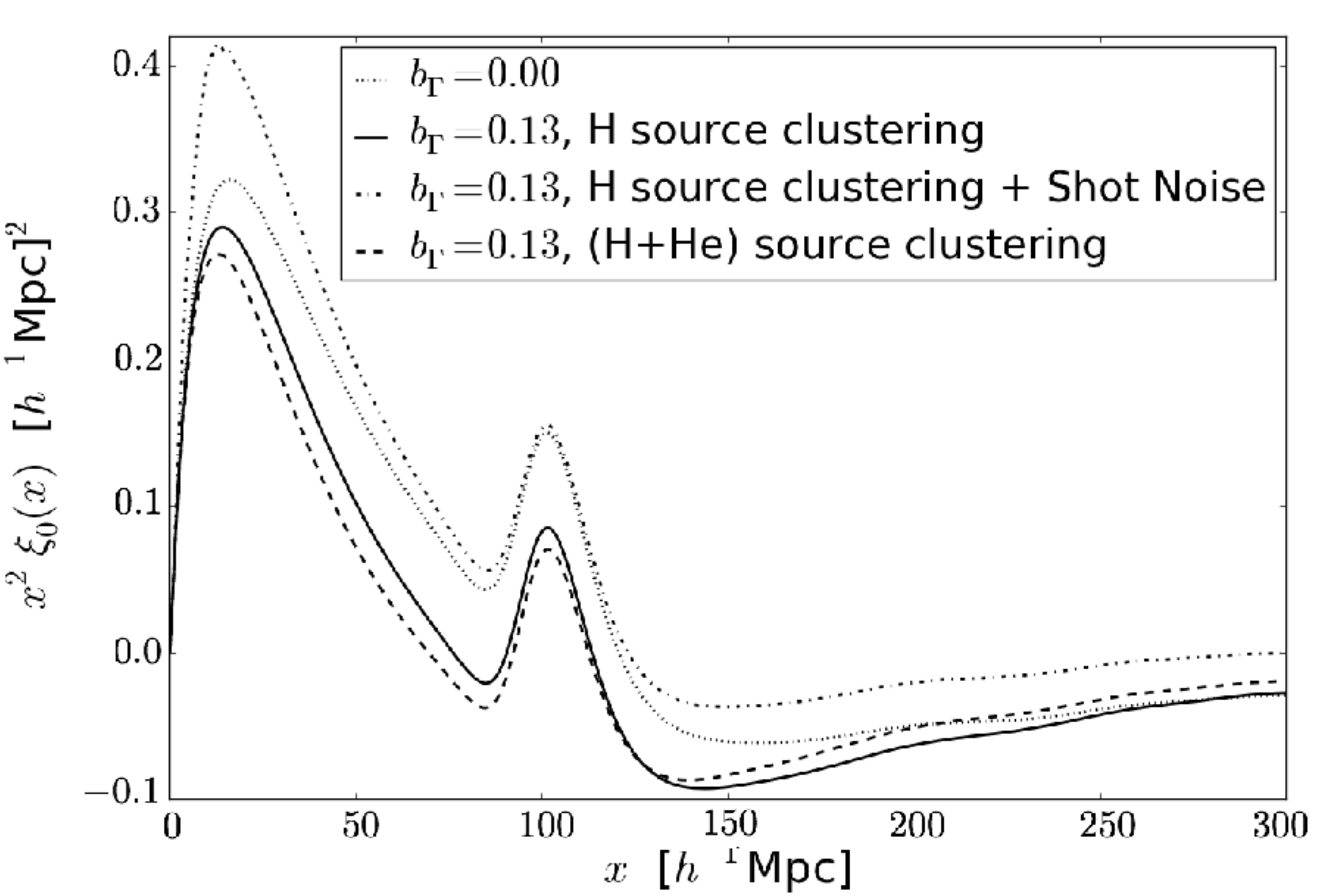}
\caption{Monopole of the \lya autocorrelation function. The dotted and
solid lines are the same as in Figures 2 and 3. The dashed line
includes the effect of helium reionisation, assuming an influence on
the gas temperature with an effective mean free path $\lambda_{0e}=
30\hmpc$. The dash-dot line shows the effect of adding the shot noise
from individual sources, multiplied by a reduction factor as explained
in the text.}\label{fig4}
\end{figure}

 A simple model for the possible helium effect on the monopole of
the correlation function,
computed as explained in section 2.4, has been included in the dashed
line in Figure \ref{fig4}. We have assumed a mean free path for the
helium reionisation influence on the gas temperature of $\lambda_{0e}=
\kappa_{0e}^{-1}=30 \hmpc$, and a relation between the gas temperature
fluctuation and helium-ionising intensity fluctuation of $b_T=0.1$,
implying $b_e=0.0084$ (see section 2.4). The small value that we
estimate for this bias factor means that the effect from the imprint
on the gas temperature that may be left from double helium reionisation
is very small, even with a much smaller mean free path than for the
case of hydrogen. However, the effects might be more substantial if
different spectra of the ionising sources in regions of different
density gave rise to a larger variation of gas temperature than our
assumed value $b_T=0.1$.

~\par The effect of shot noise is also analysed in Figure \ref{fig4}.
The dotted and solid lines are again like in Figures \ref{fig2} and
\ref{fig3} (only the monopole is shown here), and the dashed line adds
to the solid one the shot noise term from equation (\ref{eq:xishot}),
multiplied by $b_\Gamma^2$, and multiplied also by a reduction factor
that we now describe. If quasars are the sources of the ionising
background with the luminosity function used in section 2.2, and they
are isotropic and constant, the shot noise is an extremely large effect
which brings the value of the correlation function near the BAO peak, at
$x\simeq 100 \hmpc$, to $x^2 \xi_0(x)\simeq 2 (\hmpc)^2$, well above the
upper bound of the axis in Figure \ref{fig4}. However, as discussed at
the end of section 2.2, the shot noise term is likely to be reduced by
the contribution from galaxies to the ionising background (by the factor
$C/C_q$ in equation \ref{eq:norm}), the fraction of solid angle over
which quasars emit their radiation, and the fraction of the time over
which an individual quasar is emitting. For the purpose of
visualisation, we multiply the shot noise by the overall factor 
\begin{equation}
 {C\over C_q} f_\Omega f_t = {1\over 4}
 {10 \hmpc \over {\rm max}(x, 10\hmpc) } ~.
\end{equation}
The effect of variability in reducing the shot noise may reasonably be
expected to scale as $x^{-1}$, because two points in the \lya forest
are affected by the same luminosity of a certain quasar only if they are
both within the paraboloid of constant retarded time for the light
emitted by the quasar that ionises the gas producing the observed \lya
absorption, and this paraboloid has a fixed width determined by the
duration of the quasar luminous phase.
We stress, however, that the effects of anisotropy and variability
are complicated and that here we multiply the shot noise by this
simple reduction factor for display purposes.

~\par The result in Figure \ref{fig4} shows that the shot noise can be a
large and highly significant effect. Clearly, the case of
constant and isotropic quasars was already ruled out by the
observations of \cite{Slosar2011}, which showed that the correlation
function was well fitted by the linear approximation that generalises
the redshift distortion effects \citep{Kaiser1987} to the \lya forest.
Even with a large reduction of the shot noise term, the effects are
likely to be comparable to the source and absorber clustering term,
and this will make the interpretation of any observed differences
from linear theory to be complicated. One can hope, nevertheless,
that by combining a detailed observation of the monopole and quadrupole
terms, and using joint constraints from cross-correlations of the \lya
forest with quasars and other objects in addition to the \lya
autocorrelation, the impact of the biasing of sources and absorbers
and the shot noise from a complex population of ionising sources can
be disentangled in the future.

\section{Discussion and Conclusions}

~\par The first observational determination of the large-scale \lya
power spectrum in redshift space by \cite{Slosar2011} showed a
remarkably good agreement with the simple linear theory of redshift
space distortions with the Cold Dark Matter power spectrum. The same
conclusion was reached from measurements of the cross-correlations
with damped \lya systems and quasars \citep{Font12,Font13}. However,
the ionising intensity fluctuations should have an impact on these
correlations. We have presented an analytical framework in this paper
to model these effects in the \lya autocorrelation, which can also be
easily generalised to the cross-correlation with quasars or other
objects, assuming they contribute as sources of the ionising background.
Our conclusion from the results obtained in
a few illustrating cases is that both the clustering term that measures
how sources and absorbers of the ionising background trace the mass
density fluctuations, and the shot noise term that depends on the
luminosity function and other properties of the sources, have an
important and measurable effect on the monopole and quadrupole of the
\lya autocorrelation. A substantial broadband term is added as a
contamination to this autocorrelation, which is being marginalised
over in present studies that are focused on inferring the scale of
the Baryon Acoustic Oscillation peak \citep{Busca2013,Slosar2013}.
As the modeling of the spectral calibration and quasar continua and the
accuracy of the \lya correlation measurements in BOSS and upcoming
surveys improve in the future, we can look forward to a detection of the
broadband terms induced by radiation fluctuations discussed in this
paper.

~\par There are several parameters that are important in determining how
the \lya correlation is modified by intensity fluctuations. These are
the quantities appearing in equation (\ref{eq:bias}) for the effective
\lya bias factor, and the mean free path of ionising photons. The
additional shot noise term is also dependent on many characteristics
of the sources: the luminosity function, luminosity history and emission
anisotropy. Disentangling all these effects from a detailed measurement
and model fit to the redshift-space autocorrelations and
cross-correlations will probably be a difficult challenge. However,
if the emission properties and typical luminosity histories of quasars
can be well understood from an accurate determination of the quasar-\lya
cross-correlation, it should be possible to model the shot-noise
contribution to the autocorrelation and to infer from the observations
some constraints on the biasing terms that affect the source clustering
term. It is also worth noting that in the \lya power spectrum, the
term proportional to $\mu_k^4$ is affected by neither the source
clustering nor shot noise effects, and the other two terms proportional
to $\mu_k^2$ and independent of $\mu_k$ can in principle be used to
separate the influence of the source clustering and shot noise effects
(the term proportional to $\mu_k^2$ is not affected by shot noise for
constant and isotropic sources, but would acquire a contribution for
anisotropic and variable sources). The three-dimensional
\lya power spectrum therefore provides a way of separating the radiation
influences by separating the multipole terms, which are predicted to
have the specific features near the scale of the mean free path shown in
Figure \ref{pfig} can then be compared to constraints obtained from
cross-correlations.

 ~\par The conclusions of our work are in agreement with those of
Pontzen (2014), who has presented very similar ideas with a somewhat
different mathematical treatment. There are a few differences in the
way that absorbers are treated, and our incorporation of the redshift
distortion effects allows us to predict the different behavior of the
monopole and quadrupole terms in the \lya power spectrum, but the basic
conclusions of the two papers are similar.

~\par  While the radiation intensity fluctuations make the large-scale
\lya forest correlations substantially more difficult to interpret as
a tracer of the primordial fluctuations in the universe, these
complications practically do not affect the measurement of the Baryon
Acoustic Oscillation scale, and they should constitute a new motivation
for studying the evolution of the source and absorber population of
the ionising background.

\section*{Acknowledgments}

SGG thanks the APC, especially the "Cosmology \& Gravitation" group,
for their hospitality during part of the time when this work was
being carried out. We also thank David Kirkby, Pat McDonald and Andrew Pontzen
for stimulating discussions and for sharing their work on this subject.
This work is supported in part by Spanish grant AYA-2012-33938.

\label{lastpage}

\end{document}